# Printing surface charge as a new paradigm to program droplet transport

Qiangqiang Sun[1], Dehui Wang[1], Jiahui Zhang[2], Yanan Li[3], Shuji Ye[2], Jiaxi Cui[1], Longquan Chen[1], Zuankai Wang[3*], Hans-Jürgen Butt[4*], Vollmer Doris[4], Xu Deng[1*]

1. Institute of Fundamental and Frontier Sciences, University of Electronic Science and Technology of China, Chengdu, China
2. Department of Chemical Physics, Hefei National Laboratory for Physical Sciences at the Microscale, University of Science and Technology of China, Hefei, China.
3. Department of Mechanical Engineering, City University of Hong Kong, Hong Kong, China
4. Department of Physics at Interfaces, Max Planck Institute for Polymer Research, Mainz, Germany

**Directed, long-range and self-propelled transport of droplets on solid surfaces, especially on water repellent surfaces, is crucial for many applications from water harvesting to bio-analytical devices[1-9]. One appealing strategy to achieve the preferential transport is to passively control the surface wetting gradients, topological or chemical, to break the asymmetric contact line and overcome the resistance force[10-16]. Despite extensive progress, the directional droplet transport is limited to small transport velocity and short transport distance due to the fundamental trade-off: rapid transport of droplet demands a large wetting gradient, whereas long-range transport necessitates a relatively small wetting gradient. Here, we report a radically new strategy that resolves the bottleneck through the creation of an unexplored gradient in surface charge density (SCD). By leveraging on a facile droplet printing on superamphiphobic surfaces as well as the fundamental understanding of the mechanisms underpinning the creation**

**of the preferential SCD, we demonstrate the self-propulsion of droplets with a record-high velocity over an ultra-long distance without the need for additional energy input. Such a Leidenfrost-like droplet transport, manifested at ambient condition[17], is also genetic, which can occur on a variety of substrates such as flexible and vertically placed surfaces. Moreover, distinct from conventional physical and chemical gradients, the new dimension of gradient in SCD can be programmed in a rewritable fashion. We envision that our work enriches and extends our capability in the manipulation of droplet transport and would find numerous potential applications otherwise impossible[17, 18].**

When deposited on water repellent surfaces or superamphiphobic surfaces, free droplets typically exhibit random motion because of their low adhesion with the underlying substrates[3, 5, 7, 8]. In many practical applications ranging from lab on a chip to water harvest, the attainment of directional and spontaneous transport of droplets on these superamphiphobic surfaces is desired, which has been widely demonstrated by the introduction of physical or chemical gradients[1, 2, 4, 6, 9, 11]. However, the transport of droplets based on these strategies suffers from limited transport distance and low transport velocity, owing to the fundamental trade-off underpinning droplet hydrodynamics. Indeed, rapid droplet transport requires a large wetting gradient, which in turn restricts the range of transport distance[10, 12, 14, 15]. On the other hand, a long transport distance necessitates a relatively small wetting gradient, thus leading to a small transport velocity.

In contrast to our conventional perception, we observe that free droplets deposited on a special superamphiphobic surface exhibit a spontaneous, directional motion characterized by both a large transport velocity and an ultra-long transport distance (Fig. 1a, Supplementary Video 1, part 1). Such a specular transport is reminiscent of Leidenfrost phenomenon, in which droplets can migrate at a high mobility over a long distance. However, distinct from the Leidenfrost effect which occurs at high temperature resulting from the rectification of vapour flow generated from the evaporating droplets, such a Leidenfrost-like transport occurs at ambient condition.

Even when the superamphiphobic surface is placed vertically or upside down, such a Leidenfrost-like droplet transport still manifests, suggesting the stability and generality of our finding (Fig. 1b, Fig. S1 and Supplementary Video 1, part 2 and Supplementary Video 2).

How to achieve such a Leidenfrost-like transport even without the use of any additional energy input? Herein we propose an unexplored strategy that is capable of modulating the droplet dynamics through the formation of a preferential gradient in SCD on a superamphiphobic surfaces, a gradient which is totally different from previous approaches which leverage on topological or chemical gradient (Fig. 1c). Briefly, the superamphiphobic surface is constructed on a 170 μm-thick thin glass substrate coated with a 10 μm-thick nanoporous $SiO_2$ layer and hydrophobization by chemical vapor deposition of 1H,1H,2H,2H-perfluorooctyltrichlorosilane (PFOTS)[19]. We show that the preferential charge density gradient can be created by using a facile droplet impact approach. Briefly, the droplet is released at a pre-determined Weber number, defined as the ratio of the kinetic energy to the surface tension, $We = \rho d v^2 / \gamma$. Here, $\rho$, $d$, $v$ and $\gamma$ are the density, diameter, impact velocity and surface tension of the droplet, respectively[20]. Upon impact, the droplet spreads, retracts and rebounds from the superamphiphobic surface. In this process, surface charges are generated mainly at the confined to the maximum spreading diameter of the droplet as a result of the contact electrification (Fig. 2a)[21].

To validate the presence of surface charges, we further performed sum frequency generation spectroscopy (SFG) measurements to examine the ion distribution at the water/solid interface and quantified the stored charges in the bouncing droplet by using a Faraday cup[22]. When a water droplet is in contact with the fluorine groups, an electric double layer is formed at their interface[23-25]. Owing to the strong polarity of the carbon-fluorine bond, hydroxyl groups ($OH^-$) in the water droplet tend to adsorb to the superamphiphobic surface[21, 26, 27]. Thus, when the droplet detaches from the superamphiphobic surface, the surface becomes negatively charged due to trapping of the negative charges[28, 29]. The peak at 3700 $cm^{-1}$ indicates an enrichment of free OH at the solid/water interface (Fig. 2b and Supplementary Information section 2.2).

The SCD per projected area ($\rho_Q$) generated on the solid surface after the bouncing of the impinging droplet can be estimated as $\rho_Q = qA_{LS}/A_D$, where $q$ is the charge produced per unit actual contact area, $A_{LS}$ is the actual contact area, and $A_D$ is the projected liquid-solid contact area. Based on a simple hydrodynamic analysis, $A_{LS}/A_D$ is estimated as $A_{LS}/A_D \approx \frac{1}{S^2}\left[\frac{H}{2r_0}\pi r_0 + \pi r_0^2\right]$, with $S$ being the average center-to-center spacing of two neighbouring nanoscale posts, $r_0$ being the radius of posts, and $H$ being the liquid penetration depth which is intricately dependent on the Weber number (Fig. 2c and Supplementary Information section 2.1). Although it is challenging to quantitatively probe the relationship between the SCD and the Weber number, our experimental measurements indicate that the charge density is proportional to Weber number with a power exponent of 0.74 (Fig. 2d). Thus, the SCD can be regulated in on-demand manner by the control of the Weber number.

Based on the above SCD analysis, a preferential SCD gradient on the superamphiphobic surface can be established by the intricate control of the Weber number of printing droplets along the pre-determined pathway (Supplementary Video 3). For a droplet sitting on a pre-determined pathway of L, the SCD gradient $k$ is calculated as $k = (\rho_{Qs} - \rho_{Qe})/L$, where $\rho_{Qs}$ and $\rho_{Qe}$ are the charge density at the start and end points of the impacted path.

We further demonstrate that the SCD gradient-mediated droplet transport can occur on substrates with a wide range of thicknesses and dielectric constants. Note that the magnitude of SCD gradient is highly sensitive to the thickness and electrical conductivity of the superamphiphobic substrate (Fig. S3). As shown in the phase diagram (Fig. 2e), with increase in the thickness and dielectric constant of the substrate, the effective electric potential gradient becomes compromised owing to the accumulation of opposite charges resulting from the polarization of the dielectric substrate. Interestingly, unlike the chemical or morphological gradients which are difficult to change once they are created, the charge density on the superamphiphobic surface can be easily rewritten by using the ion wind, and as a result the droplet pathways can be reprogrammed in an on-demand manner (Fig. 2f).

To investigate how the attainment of SCD gradient programs the droplet motion, we continue to conduct the hydrodynamics analysis of the moving droplet. Assuming that the charge on every nanosphere-composed pillar is a point charge and the net electric force associated with a droplet can be expressed as $F_{Net} \sim (\varepsilon_r - 1)\varepsilon_0 k_l D_c$, where $\varepsilon_r$ is the relative dielectric constant of the droplet and $\varepsilon_0$ is the dielectric constant of empty space, $k_l$ is the SCD gradient imposed to the moving droplet (Fig. 3a). Note that owing to the extreme water repellency of the superamphiphobic substrate, the viscous energy dissipation is negligible. On the basis of the Newton's Second Law, the transport velocity $v$ scales as (Supplementary Information section 2.4)

$$v \sim (\varepsilon_r - 1)\varepsilon_0 k_l$$

Therefore, the droplet transport velocity can be regulated by controlling the charge gradient $k$ (Fig. 3b).

To illustrate the superior droplet transport performances enabled by the unique SCD gradient, we benchmark the unidirectional transport velocity and the transport distance ($L_t$) against those on reported surfaces (Fig. 3c). Specifically, the maximum transport velocity is measured to be 1.1 m/s, which is 20 times and 10 times higher than those on conventional superhydrophobic and hot surfaces (Leidenfrost effect), respectively[30]. The achievement of such a Leidenfrost-like droplet transport at room temperature directly eliminates the strict requirement of high temperature, making it feasible for various lab-on-a-chip applications. Moreover, we demonstrate that such a directional and spontaneous transport is also generic to liquids of different surface tension, viscosity, conductivity and dielectric constant (Fig. 3d, Supplementary table 1, 2 and 3).

The SCD gradient mediated droplet transport also manifests on surfaces with complex geometries. As shown in Fig. 4a and Fig. 4b, SCD gradient can be created on a circular arc-shaped pathway (Supplementary Video 4) as well as flexible surface (Supplementary Video 5). In both cases, a well-programmed droplet transport is achieved. More interestingly, by arranging surfaces with and without SCD gradient alternatively, a droplet can self-propel over virtually any long distance at very high velocity (Fig. 4c and 4d, Supplementary Video 6), which is otherwise impossible to achieve using conventional approaches.

In summary, we demonstrate a novel strategy that imparts unprecedented performances by taking advantage of unexplored surface charge gradient. Distinct from previous approaches that rely on the manipulation of structural or chemical gradient, the discovery of such a new transport mechanism dramatically enriches our capability to control droplet hydrodynamics. From a broader perspective, the generality of our approach will open a new avenue for a wide range of applications ranging from microfluidics, to energy, physical and biotechnology applications.

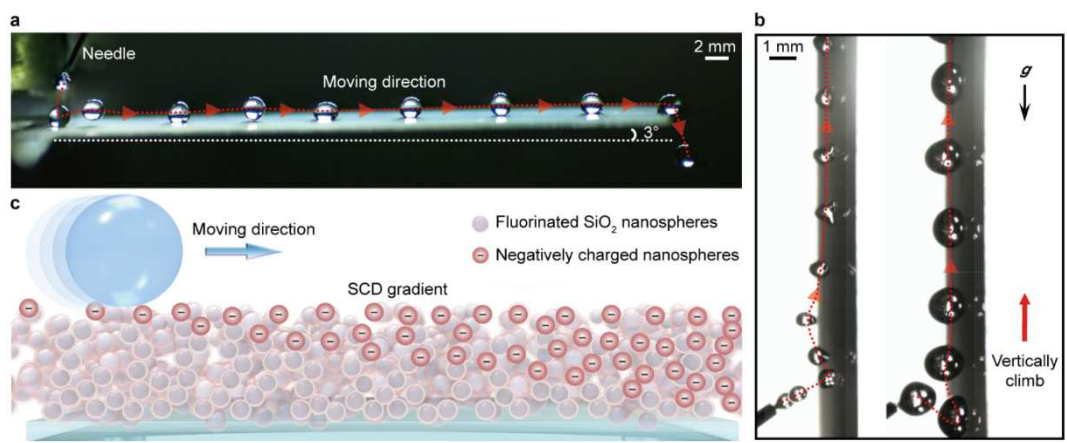

**Figure 1 | Droplet transport mediated by printable surface charge density gradient. a,** The time-lapse trajectory of a water droplet on a superamphiphobic surface decorated with SCD gradient at a climbing angle of 3°. **b,** The time-lapse trajectory of droplets with different radii moving upward on a vertically placed superamphiphobic surface decorated with SCD gradient. **c,** Schematic of droplet self-propulsion on superamphiphobic surface decorated with SCD gradient.

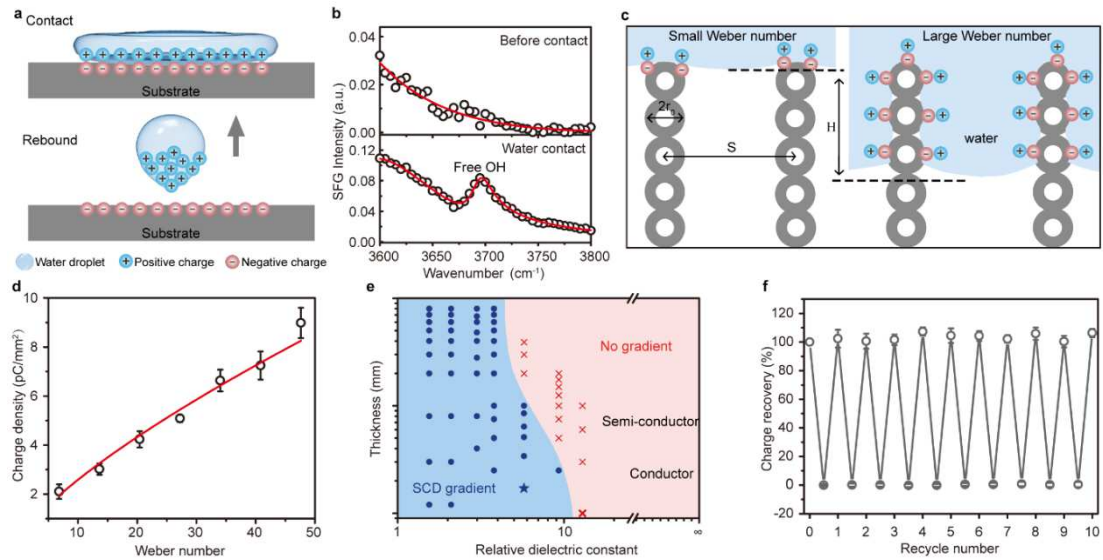

**Figure 2 | Charge characterization and charge density gradient formation. a,** Schematic of charging a superamphiphobic surface by droplet impact. **b,** Sum frequency generation spectra of a fluorosilane surface taken before and during contact with water. **c,** Schematic of the position of the air/water/solid three-phase contact line for different impact pressures. **d,** Plot of the charge density of the superamphiphobic surface (substrate: 170 μm-thick glass) as a function of Weber numbers (0.1 m/s ≤ v ≤ 0.7 m/s). **e,** Effective gradient phase diagram of a droplet with the substrates as a function of the thickness and dielectric constant of substrates. Superamphiphobic coating on different substrates and charged by droplet impact ($We$ = 48) (x-axis shows the dielectric constant, y-axis shows the thickness of the substrate). An 8 μl water droplet was released with 5 mm away from the charge density gradient region. Blue: droplets self-propelled and moved towards the charge density gradient region after released. Light red: droplets remain at rest after released. The blue pentastar represents the substrate (G-170) used in our experiment. **f,** Re-writability of the charge gradient on the superamphiphobic surface. The charge gradient can be repeatedly created by droplet impact and removed by ion wind.

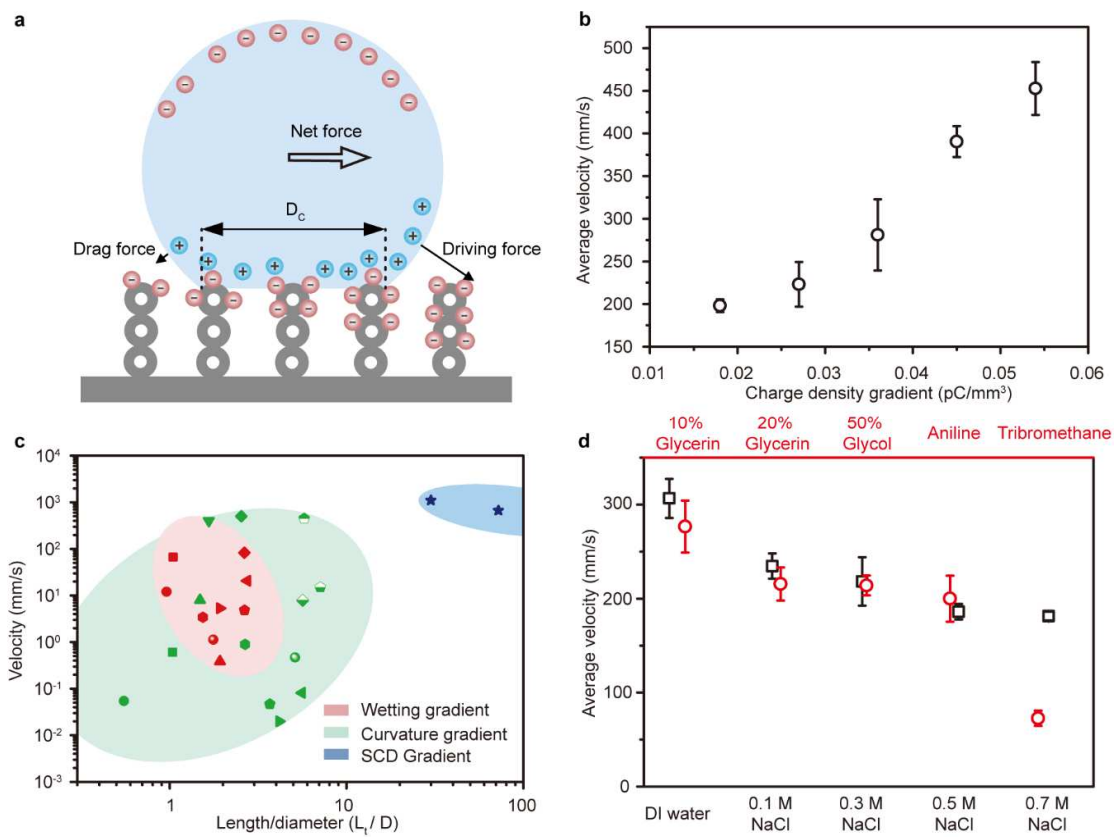

**Figure 3 | Self-propulsion mechanism and performance control. a,** Schematic of droplet transport mechanism. Because of the unbalanced electric force induced by SCD gradient, the droplet self-propels towards the direction with a larger SCD gradient. **b,** Transport velocity as a function of the SCD gradient $k$. **c,** Comparison of the droplet transport performances among different surfaces with a surface free energy gradient. The red and green areas denote the reported velocities as a function of the transport length per droplet diameter ($L_t/D$) for surfaces with wettability gradients and asymmetric geometries, respectively (Fig. S4). The blue star represents our designed superamphiphobic surface with a charge density gradient. **d,** Generality of droplet transportation for various liquids. The surface with SCD gradient shows wide versatility for transporting liquids from low surface tension to low dielectric constant and salt solution.

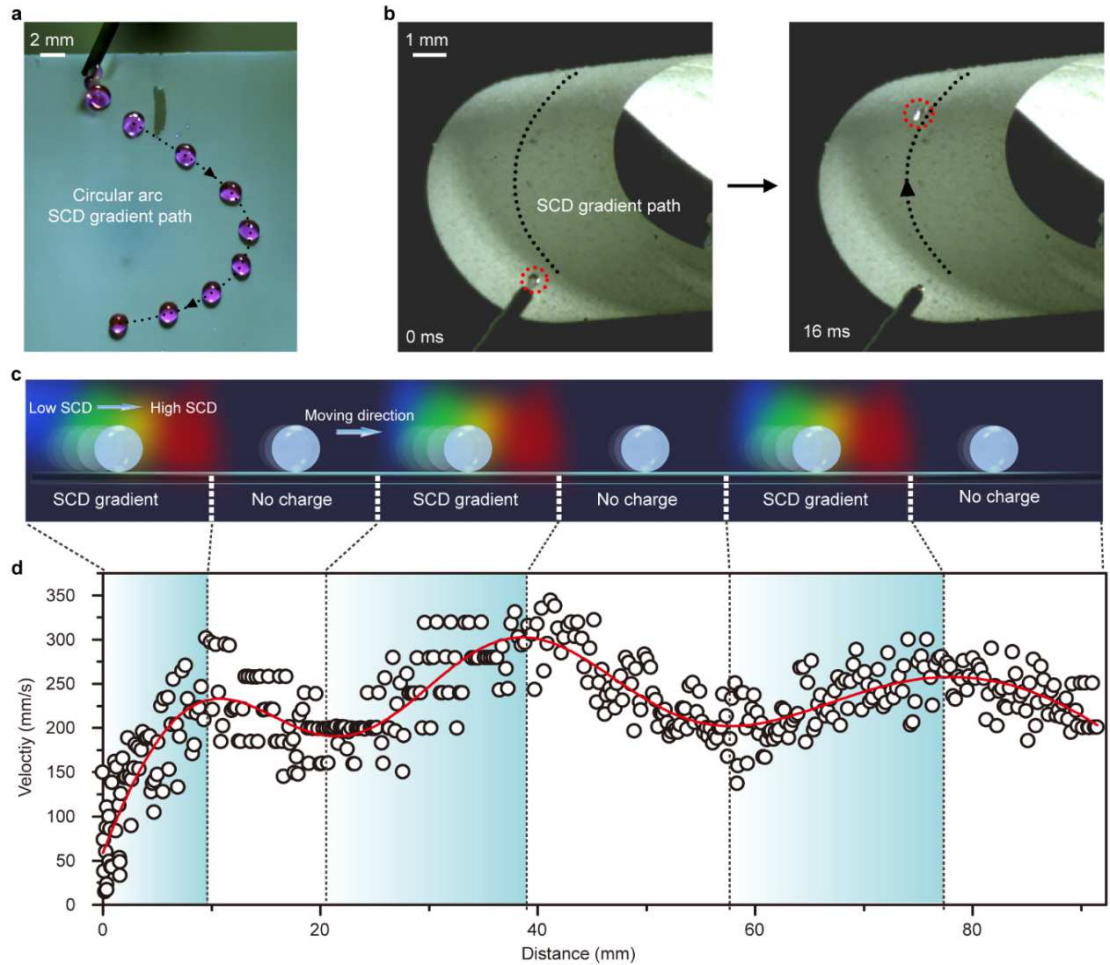

**Figure 4 | Programmable transportation. a,** Circular arc motion guided by circular arc SCD gradient path. **b,** Transport on flexible surfaces with SCD gradient. The surface was made by coating a commercial superamphiphobic coating, Ultra-Ever Dry, onto a 50 μm polytetrafluoroethylene film. **c,** Sketch of the designed surface consisting of patches with and without SCD gradient alternatively for ultra-long-range droplet transport. **d,** The variation of droplet transport velocity as a function of spatial location along the transport pathway (Supplementary Video 5).

**Supplementary Information** is available in the online version of the paper.

**Acknowledgements** This work was supported by the National Natural Science Foundation of China (21603026) and supported by Max-Planck-Gesellschaft (Max Planck Partner Group UESTC-MPIP) and the ERC advanced grant 340391-SUPRO. We thank SJ. Lin for assistance with adhesion force measurements; L.Zhou and TH.Zhang for assistance with the analytical model. S.Sun and HL.Liu for discussions.

**Author Contributions** Q.S. and X.D. conceived the research and designed the experiments. X.D., Z.W., H.J.B supervised the research. Q.S., D.W., J.Z. carried out the experiment. Q.S. and Y. Li. built the analytical models. All authors analyzed data. Q.S., S.Y., L.C., J.C., V. D. interpreted data. Q.S., X.D., Z.W., H.J.B wrote the paper.

**Author Information** Reprints and permissions information is available at www.nature.com/reprints. The authors declare no competing financial interests. Readers are welcome to comment on the online version of the paper. Correspondence and requests for materials should be addressed to X.D. (dengxu@uestc.edu.cn), W. Z. (zuanwang@cityu.edu.hk), H.J (butt@mpip-mainz.mpg.de)